*Original Article*

# The Effects of COVID-19 and the Russia-Ukraine War on Inward Foreign Direct Investment

[1]**Mohammad Saddam Hosen**, [2]**Sheikh Murad Hossain**, [3]**Md. Nahid Mia**, [4]**MR Chowdhury**

[1]*M. Phil (Researcher), Department of Management Studies, National University, Bangladesh.*
[2]*Department of Strategy and Management, University of Bedfordshire, United Kingdom.*
[3]*Assistant Professor, Department of Management Studies, National University, Bangladesh.*
[4]*School of Justice, Security and Sustainability, Staffordshire University, United Kingdom.*

***Abstract:*** *Inward Foreign Direct Investment (IFDI) into Europe and Asian developing countries like Bangladesh is experimentally examined in this study. IFDI in emerging markets has been boosted by global investment and inflow influenced by resource availability and public policy. The economic policy uncertainty on IFDI in 13 countries is explored at a time when the crisis between Russia and Ukraine war is having a global impact. Microeconomic factors affected Gross Domestic Product (GDP) growth, inflation, interest rates, and the currency rate fluctuated with IFDI, which mostly shocked during COVID-19 and the Russia-Ukraine war. With data from the World Bank and the United Nations Conference on Trade and Development (UNCTAD) database, we compile a panel dataset covering 2018-2022. The researchers used a mixture of panel and linear regression analysis using a random effect model. Our findings show that the impact of global rates hurts IFDI in 13 selected countries. There is a correlation between a country's ability to enforce contracts and the amount of Inward FDI it receives. Using the top 13 hosts of incoming FDI flows COVID-19 and Russia-Ukraine wartime series analysis gives valuable information for policymakers in the remaining countries chosen to attract IFDI inflows.*

## I. INTRODUCTION

Scholarly interest in inward Foreign Direct Investment (FDI) has grown significantly, with researchers concentrating on determining its causes and assessing its possible effects on the economies in both host and home nations. The stock of foreign direct investment is the total amount of debt and equity financing that foreign investors give to businesses that are operated at home. When a company from one nation buys another company from another nation or establishes operations in a new nation, this is known as Foreign Direct Investment (FDI). IFDI strengthens domestic economies to the inside out by boosting GDP, creating jobs, and developing infrastructure. (Grosse and Trevino, 1996). The economic crisis and the Russia-Ukraine war have changed the global FDI picture, as well as the impact on the country's local investment, especially in inflation and GDP (Charaia, Lashkhi and Lashkhi, 2022). FDI flows to the transition economies of South Asian countries, significantly Bangladesh, were impacted harder than markets in most other regions by the long-lasting adverse effects of COVID-19 (Zaman, 2023). When the COVID-19 outbreak struck in 2020, causing significant economic upheaval, the invasion of Russia-Ukraine was also seen as the most critical geopolitical calamity, and Alam et al. (2022) claimed several worldwide shocks of inward FDI as a percentage of GDP skyrocketed. Inward Foreign Direct Investment (IFDI) positively affects the introduction of novel techniques and knowledge in developing nations (Hosen, 2022; Nawo and Njangang, 2022).

The connection between IFDI flows and a number of economic indicators, such as GDP, inflation, interest rates, and exchange rates, is investigated in this research. Thirteen nations that have significantly increased foreign direct investment are the subject of the study. Annual aggregate-level data from 2018 to 2022 are used in the study. The results show that the lagged inward FDI flow factors have significant negative effects. As seen from the perspective of the host country, this is "inward investment". (Borensztein, De Gregorio and Lee, 1998). It is widely believed that inflows of FDI and other forms of international capital play a crucial influence on a country's economic development and prosperity (Burlea-Schiopoiu, Brostescu and Popescu, 2023). The impact of the Russia-Ukraine conflict on inward foreign direct investment (FDI) flow is a hotly debated topic. Other factors that may impact the nation's place in the global economy and its involvement in international economic activities. It is unable to account for long-term variations in FDI inflows following COVID-19. Given their diminished variability, this category of determinants is typically studied in relation to the investment stock. IFDI inflows may also have an impact on banking operations and monetary policy decisions. The spatial effect of FDI provides novel evidence of the relationship between IFDI flow and GDP, Inflation rate and Exchange rate, filling a significant research void in the literature. It employed spatial econometric techniques (Nurmakhanova et al., 2023) to take into consideration the direct and indirect effects in 13 nations between 2018 and 2022. According to the literature related to the Russia-Ukraine war and the COVID-19 impact on inward FDI in the case of 13 countries, including Bangladesh, there are more challenges and factors affecting the local economy.

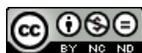

## II. LITERATURE REVIEW

As per the World Trade Organization (WTO), the global economic situation has deteriorated since the commencement of hostilities between Russia and Ukraine on February 24, 2022. Trade expansion is anticipated to decline from 4.7% to below 3.4% (WTO, 2023). The effects of the COVID-19 epidemic have severely impacted FDI in developing and emergent economies. World Economic Forum reported that 30% and 40% of worldwide FDI inflows would decrease from 2020 -2021 (WEF, 2022). Less developed and developing nations are disproportionately affected by the decline in FDI inflows.

Furthermore, Huang, Teng, and Tsai (2010) stated that the benefits of IFDI on cost reductions, increased productivity, and eventually increased production have the capacity to equalize or even outweigh the drawbacks of IFDI on distribution. Based on the existing status of middle-income countries within the global FDI market and the global value chain (Hajian Heidary, 2022), the contention remains that IFDI is more prone to adversely impacting domestic income distribution (Moosa and Merza, 2022). The reason technology overload is not taken into account is not because they are not important; rather, it is because there is no reliable indicator to determine their independent impact. (Teng et al., 2023).

Chattopadhyay et al. (2022) strong external influences of exports in the non-exports sector and higher marginal effectiveness in the exports sector compared to the non-exports sector had an important effect on exports and GDP growth. IFDI and GDP have a strong connection. (Ahmad et al., 2021). The inflation rate increased after COVID-19, raising production costs in the Russia-Ukraine war (Liadze et al., 2023). Some Asian developing countries experienced a negative impact during the Russia-Ukraine war on money exchange rate unrest in the global market (Izzeldin et al., 2023). High inflation reduces the actual value of earnings for inward-investing businesses in the nation's currency. On the other hand, a low inflation rate attracts foreign direct investment and signifies a state of internal financial stability within a country. According to Klein and Rosengren's (1994) study, two primary routes exist (Khan and Ahmad, 2021): a system dynamics model of the relationship between the COVID-19 pandemic and foreign direct investment in the worldwide supply chain and which foreign exchange rates affect FDI. (Hajian Heidary, 2022) and the relative production cost channel. The wealth effect shows that foreign investors' overall wealth increases as a result of depreciating currencies when compared to domestic investors. The depreciation of the host country's currency lowers the cost of manufacturing inputs, such as labor, land, machines, and wealth, from the standpoint of foreign investors who assess capital in foreign currency.

The study by Boateng et al. (2015) showed that, among seven advanced nations, interest rates play a critical role in determining which host countries attract Foreign Direct Investment (FDI). Uddin et al. (2019) have also revealed similar outcomes, emphasizing how the decision-making process of Korean multinational enterprises (MNEs) when choosing the best places for their factories within EU member states is greatly influenced by the low-interest rates in those countries. Ramb and Weichenrieder (2005) have presented evidence supporting the notion that host nations' low interest rates significantly attract IFDI. Albulescu and Ionescu (2018) add weight to this debate by stating that low-interest rates give investors a cost advantage. Alguacil, Cuadros, and Orts (2008), on the other hand, proposed that higher interest rates (Udomkerdmongkol, Morrissey and Görg, 2009) through the creation of profitable investment opportunities within the recipient country increase the allure of foreign investments.

As a result, a number of empirical research reviews have looked into the relationship between GDP and FDI. Chowdhury and Mavrotas' (2006) study found that, in the context of the BRICS, there was a favourable association between the growth of the host country's market size, as indicated by a higher GDP, and a rise in FDI inflows. (Chattopadhyay et al., 2022). Moreover, this research offers more proof to back up the idea that GDP plays a big role in drawing FDI over a long time horizon. A thorough analysis is required due to the multifaceted nature of the influence of IFDI and outward migration on the upgrading of global value chains (Amendolagine et al., 2019). Jude and Silaghi (2016) stated that during the past few decades, member states of the European Union (EU) have all shown a consistent upward trend in the percentage of foreign direct investment (FDI) to GDP. This pattern provides evidence for the theory of increased economic integration in the EU. Khan and Ahmad's (2021) study examines the enduring association between FDI inflows as a percentage of GDP, uncertainty in monetary policy, and the stability of the banking industry in a group of sixteen countries within the European Union (EU) between 2001 and 2015 (Kottaridi, Louloudi and Karkalakos, 2019). According to recent assessments, the Russia-Ukraine conflict has negatively influenced the world economy, stock market, energy market, commodity prices and resources.

Driffield and Karoglou (2019) found that the effect of Brexit on foreign direct investment (FDI) in the UK market varies based on the investment's motivations. In particular, Brexit is predicted to have less of an impact on foreign direct investment (FDI) in industries like retail, construction, power, and transportation that are looking for market opportunities in the UK than on FDI motivated by other factors. The attractiveness surveys by Bailey, Driffield, and Kispeter (2019) provide the information needed to perform a worldwide analysis of the number of FDI inflow projects and related jobs. Following the Russia-Ukraine war cycles and the downturn in foreign investment, the United Kingdom saw a peak in the amount of IFDI efforts in 2023.

**Table 1: IFDI Stock in the World**

| Region/economy | 2000 | 2010 | 2021 | 2022 |
|---|---|---|---|---|
| World | 7 377 201 | 19 855 669 | 47 079 311 | 44 252 759 |
| Developed economies | 5 860 038 | 13 788 303 | 32 816 197 | 29 093 016 |
| Europe | 2 491 244 | 8 381 352 | 16719061 | 15604 111 |
| European Union | 1 882 785 | 5 902 591 | 12 098 672 | 11 170 459 |
| Developing economies | 1 517 163 | 6067365 | 14263114 | 15 159744 |
| Asia | 1 023690 | 3879019 | 10846116 | 11 495416 |
| L. America and the Caribbean | 338 557 | 1 550 229 | 2 355 235 | 2 580 077 |
| Central America | 139 768 | 417 113 | 763 963 | 829 908 |
| Oceania | 1 854 | 14 694 | 30 785 | 31 724 |

*Source: UNCTAD (2023)*

UNCTAD (2023) reports that COVID-19 has an inverse relationship with foreign direct investment (FDI); Orhun (2021) found similar results regarding the impact of the Russia-Ukraine war on IFDI. According to Xia and Liu (2021), lockdowns and similar measures have a considerably worse effect on economies and IFDI activity than the real number of COVID-19-related illnesses and fatalities. Studies such as the one conducted by Strange (2020) also consider the true damage caused by the COVID-19 pandemic.

### III. METHODOLOGY

The magnitude and direction of the association between IFDI and the macroeconomic factors of GDP growth, GDP net inflow, inflation, exchange rate, and interest rate were examined and reported in this case study. Therefore, adhered to both descriptive as well as analytical methodologies. The last five years' worth of country data, including those from China, the United Kingdom, Sweden, Norway, Poland, Japan, Switzerland, Germany, Italy, Turkiye, the Netherlands, Belgium, and Bangladesh, was gathered from the WB and UNCTD database repository. This time frame provides limits on time for the study and is adequate for tracking changes in macroeconomic factors. The sample included a range of IFDI-related industries.

**Figure 1: Study area map (country selected by the dark line)**

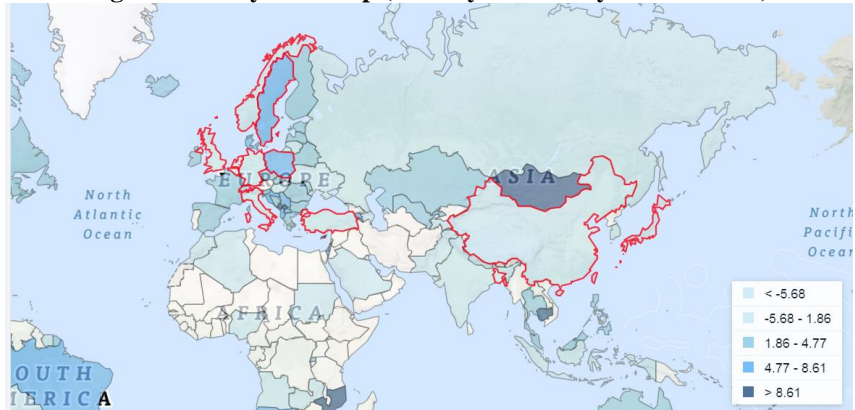

*Source: World Bank, (2023)*

The WB, UNCTD, and IMF reports of datasets for 2018–2022 were the only sources of data used in the study. As a result, secondary data were mostly used in this study. The macroeconomic data came from secondary sources and were taken from economic surveys conducted between 2018 and 2021. Previous research and theoretical frameworks suggested by capital structure concepts served as the basis for the model and variables used in this investigation. The following provides an explanation of the model used in the analysis:

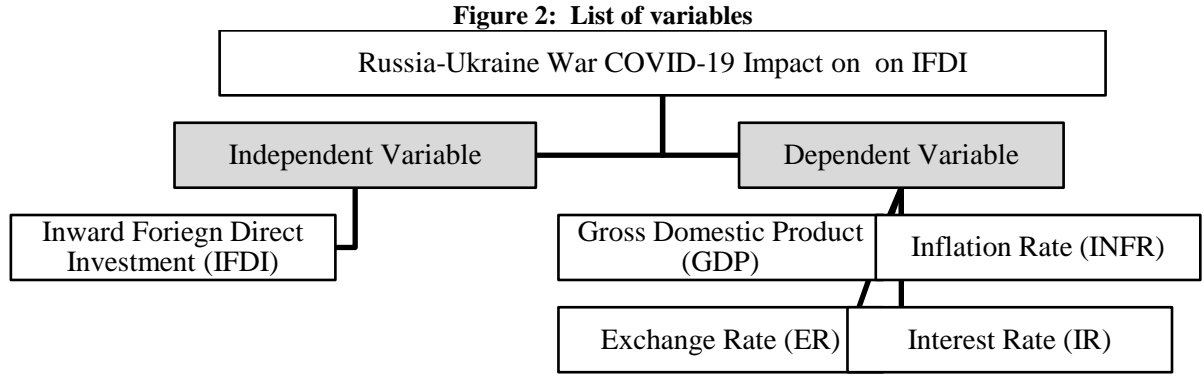

**Figure 2: List of variables**

*Source: Illustrated by Author (2023)*

**Table 2: Variable Predicted Hypothesis Details**

| Control Variable | Database | Hypothesized Result |
|---|---|---|
| GDP (Annual Growth %) | UNCTAD | $H_1$: Positive GDP and significant IFDI influence in selected countries. |
| Inflation Rate | WB, IMF | $H_2$: Positive inflation rates considerably impact IFDI in selected countries. |
| Exchange Rate | WB | $H_3$: The exchange rate is negative and significantly influences the IFDI in selected countries. |
| Interest Rate | UNCTAD, OECD | $H_4$: Interest is negative and significantly influences the IFDI in selected countries. |

**Table 3: Five Years Inward FDI Stock of Selected Country (2018-2022)**
*USD Million*

| Selected Country | 2018 | 2019 | 2020 | 2021 | 2022 |
|---|---|---|---|---|---|
| Belgium | 502433.4 | 583746.5 | 604255.5 | 555736.1 | 523854.7 |
| Germany | 940576.8 | 963565.9 | 1153099 | 1057990 | 1007533 |
| Italy | 434643.4 | 443554.2 | 490196.7 | 449961.7 | 448,493 |
| Netherlands | 1458721 | 1427625 | 2721329 | 2744450 | 2683600 |
| Poland | 229,527 | 240586.4 | 256007.8 | 270718.6 | 269840.1 |
| Sweden | 314353.1 | 326633.9 | 395715.4 | 387482.9 | 353790.7 |
| Norway | 155236.8 | 170541.7 | 167096.1 | 211,593 | 145,513 |
| Switzerland | 1075443 | 1100354 | 1183255 | 1038359 | 1036890 |
| United Kingdom | 1996777 | 2152764 | 2656647 | 2689966 | 2698563 |
| Japan | 204523.6 | 223809.7 | 250070.1 | 241125.5 | 225367.1 |
| China | 1628261 | 1769486 | 1918828 | 3633317 | 3822449 |
| Bangladesh | 17061.63 | 17784.98 | 19394.76 | 21581.86 | 21158.2 |
| Turkiye | 145302 | 160648 | 229961 | 139970 | 164909 |

*Source: World Bank, (2023a)*

**Table 4: Country-wise GDP growth (2018-2022)**

| Selected Country | 2018 | 2019 | 2020 | 2021 | 2022 |
|---|---|---|---|---|---|
| United Kingdom | 1.7 | 1.6 | -11 | 7.6 | 4.1 |
| Turkiye | 3 | 0.8 | 1.9 | 11.4 | 5.6 |
| Switzerland | 2.9 | 1.1 | -2.4 | 4.2 | 2.1 |
| Sweden | 2 | 2 | -2.2 | 5.4 | 2.6 |
| Poland | 5.9 | 4.5 | -2 | 6.8 | 4.9 |
| Norway | 0.8 | 1.1 | -1.3 | 3.9 | 3.3 |
| Netherlands | 2.4 | 2 | -3.9 | 4.9 | 4.5 |
| Japan | 0.6 | -0.4 | -4.3 | 2.1 | 1 |
| Italy | 0.9 | 0.5 | -9 | 7 | 3.7 |
| Germany | 1 | 1.1 | -3.7 | 2.6 | 1.8 |
| China | 6.7 | 6 | 2.2 | 8.4 | 3 |
| Belgium | 1.8 | 2.3 | -5.4 | 6.3 | 3.2 |
| Bangladesh | 7.3 | 7.9 | 3.4 | 6.9 | 7.1 |

*Source: UNCTAD, (2023a)*

**Table 5: Country-wise Inflation Rate (2018-2022)**

| Selected Country | 2018 | 2019 | 2020 | 2021 | 2022 |
|---|---|---|---|---|---|
| United Kingdom | 1.7 | 2.1 | 5.9 | 0 | 5.4 |
| Turkiye | 16.5 | 13.8 | 14.9 | 29 | 96.1 |
| Switzerland | 0.8 | -0.1 | -0.7 | 1.1 | 3.3 |
| Sweden | 2.4 | 2.5 | 2 | 2.9 | 5.7 |
| Poland | 1.2 | 3 | 4.3 | 5.1 | 11.5 |
| Norway | 6.7 | -0.5 | -2.5 | 17.1 | 28 |
| Netherlands | 2.4 | 3 | 1.9 | 2.4 | 5.3 |
| Japan | 0 | 0.6 | 0.9 | -0.2 | 0.2 |
| Italy | 1.1 | 0.9 | 1.6 | 0.6 | 3 |
| Germany | 2 | 2.1 | 1.8 | 3.1 | 5.5 |
| China | 3.5 | 1.3 | 0.5 | 4.6 | 2.2 |
| Belgium | 1.5 | 1.7 | 1.5 | 2.8 | 5.9 |
| Bangladesh | 5.8 | 3.7 | 3.8 | 4.1 | 5 |

***Source:** IMF, (2023)*

**Table 6: Real effective exchange rate index (2018-2022)**

| Selected Country | 2018 | 2019 | 2020 | 2021 | 2022 |
|---|---|---|---|---|---|
| United Kingdom | 99 | 98.6 | 98.8 | 102.6 | 101.2 |
| Switzerland | 103.2 | 104.2 | 108.2 | 105.5 | 105.8 |
| Netherlands | 99.8 | 99.8 | 101.9 | 102.1 | 102.2 |
| Turkiye | 111.82 | 112.44 | 92.37 | 70.87 | 83.61 |
| Germany | 97 | 95.5 | 96.4 | 97.1 | 93.6 |
| China | 122 | 121.2 | 123.6 | 127.3 | 125.8 |
| Italy | 96.9 | 94.6 | 95.1 | 94.9 | 93 |
| Japan | 74.5 | 76.6 | 77.3 | 70.7 | 61 |
| Norway | 86 | 83.6 | 78.1 | 83.1 | 80.8 |
| Poland | 93.5 | 92.4 | 92.9 | 92.6 | 93.9 |
| Sweden | 89.3 | 85.7 | 87.7 | 90.3 | 84.8 |
| Belgium | 101.3 | 99.8 | 101.2 | 101.4 | 101 |
| Bangladesh | 100.5 | 106.4 | 113.5 | 110.7 | 111.3 |

***Source:** OECD, (2023a)*

**Table 7: Interest Rate (2018-2021)**

| Selected Country | 2018 | 2019 | 2020 | 2021 | 2022 |
|---|---|---|---|---|---|
| United Kingdom | 6.9 | 6.1 | 4.3 | 6.5 | 5.25 |
| Switzerland | 1 | 0.9 | 0.7 | 0.6 | 1.75 |
| Netherlands | 2.3 | 2 | 1.6 | 1.4 | 4.3 |
| Turkiye | 8.3 | 6.9 | 9 | 10.2 | 25 |
| Germany | 1.8 | 1.5 | 1 | 0.9 | 4.25 |
| China | 4.5 | 4.7 | 4.3 | 3.75 | 3.75 |
| Italy | 8.4 | 7.8 | 6.7 | 7.2 | 4.3 |
| Japan | 8.9 | 8.5 | 5.9 | 6.2 | 1 |
| Norway | 0.9 | 0.8 | 0.7 | 0.6 | 3.75 |
| Poland | 4 | 3.7 | 3 | 2.8 | 6.75 |
| Sweden | 1.3 | 1.1 | 0.7 | 0.6 | 3.75 |
| Belgium | 4.7 | 4.4 | 3.9 | 3.5 | 0 |
| Bangladesh | 21.1 | 20.5 | 22.6 | 24 | 6 |

***Source:** World Bank, (2023b)*

Only listwise marginalization is allowed for contrast descriptive statistics, which indicates that all dependent and independent variables have acceptable values for a given table.

## IV. RESULTS AND DISCUSSION
**Table 8: Descriptive statistics on the independent and dependent variable**

| Variable | Mean | Median | Mode |
|---|---|---|---|
| Inward FDI | 2310.906 | 2284.055 | 1292.56[a] |
| GDP Growth | 51206.093 | 46417.580 | 27771.89[a] |
| Inflation Rate | 2179.711 | 2168.750 | 846.55[a] |
| Exchange Rate | 34954.771 | 35636.880 | 25319.16[a] |
| Interest Rate | 81.481 | 81.250 | 77.750[a] |

*Source: SPSS V. 26*

Table 8 displays the calculated values from 5 years of data in the study, which is 2310.906 mean with mode of 1292.56 on foreign direct investment. The GDP growth 51206.093 means shows that, on 46417.58 medians, domestic investment significantly impacted the 2179.711 mean, which affected the 846.55 mode. The export category for a subset of 13 countries, including the United Kingdom, had a mean score of 34954.77 for the IFDI inflow rate and a high score median of 35636.880 for the years 2008–2022. The mean foreign exchange rate is lower at 81.481, with medians of 81.250 and a mode score of 126.65 in relation to IFDI.

The Pearson Correlation value, shown over in the square red box in this instance, is a 0.95 significance value best fitted. Pearson's r varies between +1 and -1, where +1 is a perfect positive correlation, and -1 is a perfect negative correlation. 0 means there is no linear correlation at all.

**Table 9: Two-tailed Pearson correlation**

| Variable | | IFDI | GDP | Inflation Rate | Exchange Rate | Interest Rate |
|---|---|---|---|---|---|---|
| IFDI | Pearson Corr. | 1 | .718* | .764* | .819** | .744* |
| | Sig. | | .019 | .010 | .004 | .014 |
| GDP Growth | Pearson Corr. | .718* | 1 | .977** | .773** | .898** |
| | Sig. | .019 | | .000 | .009 | .000 |
| Inflation Rate | Pearson Corr. | .764* | .977** | 1 | .803** | .947** |
| | Sig. | .010 | .000 | | .005 | .000 |
| Exchange Rate | Pearson Corr. | .819** | .773** | .803** | 1 | .669* |
| | Sig. | .004 | .009 | .005 | | .034 |
| Interest Rate | Pearson Corr. | .744* | .898** | .947** | .669* | 1 |
| | Sig. | .014 | .000 | .000 | .034 | |
| | Sig. | .342 | .084 | .090 | .402 | .065 |
| *. Correlation is significant at the 0.05 level (2-tailed). | | | | | | |
| **. Correlation is significant at the 0.01 level (2-tailed). | | | | | | |
| Listwise N= 13 countries | | | | | | |

With all variables 100% associated and one uninterested relationship in Table 9, there is a substantial positive relationship between IFDI and the dependent variable. This relationship is significant as the significance value is the most powerful, which is greater compared to the typical significance level of 0.95 percent. IFDI and GDP have a significant positive correlation; their significance value is.0718, higher than the conventional significance level of 0.05, indicating that the relationship is meaningful. There is a 71% correlation between the two variables. Because the significance value in Table 9 is 1, which is higher than the conventional 95% significance level, it is evident that there is a strong relationship between INFR and IFDI. There is a 100% correlation between the two variables and IFDI. Because the two variables have a correlation of 0.669, which is better compared to the conventional significance level of 66%, the ER linked to inward foreign direct investment concentrates on the typical relationship. Because the significance value of IR on FDI evaluates is 0.744, it exhibits a better correlation than the conventional significance level of 74% for both variables.

One reliable technique for identifying the variables that influence a foreign direct investment (FDI) is regression analysis. It is possible to ascertain with complete accuracy which variables are the most significant, which factors are negotiable, and which factors affect each other through the process of performing a regression.

**Table 10: Regression model summary**

| R | R Square | Adjusted R Square | Std. Error of the Estimate |
|---|---|---|---|
| .880a | .794 | .491 | 3.30777 |
| a. Predictors: (Constant), GDP, INFR, ER, IR | | | |
| b. Dependent Variable: IFDI | | | |

The strength of the correlation is measured by the square of the multiple relationship coefficients in Table 10. When $R^2$ is larger, then the correlations between the independent variables are low. Here, the adjusted R square is 79%, which means the independent variables can explain 77% of the total variance of IFDI with the standard error of 3.30%.

**Table 11: Analysis of variance (ANOVA)**

| Mode | Sum of Squares | df | Mean Square | F | Sig. |
|---|---|---|---|---|---|
| Regression | 3537618.227 | 5 | 707523.645 | 2.738 | .175b |
| Residual | 1033507.168 | 4 | 258376.792 | | |
| Total | 4571125.396 | 9 | | | |
| a. Dependent Variable: IFDI | | | | | |
| b. Predictors: (Constant), GDP Growth, INFR, ER, IR | | | | | |

Table 11 tests the null hypothesis to determine the validity of the model, which is set to determine the factors affecting IFDI. If the level of significance is less than .005, then the null hypothesis is rejected, and the model is significant. Here, we can see from the ANOVA table that the significance level is .000. So, we can say that the null hypothesis is rejected and the model is significant. Table 11 shows that the total mean of square 707523.645 is between 258376.792 of IFDI of 2018-2022.

**Table 12: Regression coefficients**

| Variable | Unstandardized Coefficients | | Standardized Coefficients | t | Sig. |
|---|---|---|---|---|---|
| | B | Std. Error | Beta | | |
| (Constant) | -18884.755 | 17730.780 | | -1.065 | .747 |
| GDP Growth | .009 | .047 | .246 | .188 | .860 |
| INFR | -.679 | 1.327 | -1.085 | -.512 | .636 |
| ER | .101 | .060 | .826 | 1.692 | .166 |
| IR | 228.026 | 216.733 | 1.037 | 1.052 | .352 |
| Dependent Variable: IFDI | | | | | |

By the coefficients table, we can drive the following equation:
Inward Foreign Direct Investment (IFDI) = $0.747 + 0.860 * \alpha + 0.636 * \beta + 0.166 * \gamma + 0.352 * \zeta$
$\alpha$ = Gross Domestic Product (GDP)
$\beta$ = Inflation Rate (INFR)
$\gamma$ = Exchange Rate (ER)
$\zeta$ = Interest Rate (IR)

According to the equation, if α is changed by one unit and all other variables remain constant, then I will increase by 0.860 units. IFDI will increase by 0.636 units for every unit change in variable β if all other variables are constant or under authority. IFDI will increase by 0.166 units for every unit change within variable γ if all other variables are constant or under control. IFDI will increase by 0.352 units for every unit change in variable ζ if all other variables are constant or under control if every variable is an entity of authority.

**Figure 3: Normal Probability Plot**

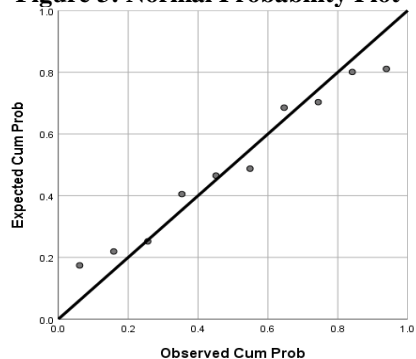

The global inflow of foreign direct investment significantly surged in 2018, with a particular emphasis on emerging countries (Figure 3). Following 2022, many developing countries—Bangladesh among them—took the initiative to remove restrictions and implement bilateral investment laws to promote trade and global direct investment. As a result, there was a noticeable uptick in IFDI, business activity, and investment growth in a number of developed countries, with a focus on the UK. There has been a discernible increase in the percentage of inward foreign direct investment (FDI) relative to GDP in the Asian region between 2018 and 2022. The impact of IFDI on a panel dataset comprising 13 nations with higher inflation rates following COVID-19 and the Russia-Ukraine war is greatly enhanced by this study (Table 8). The UK, Turkey, Switzerland, Sweden, Poland, Norway, the Netherlands, Japan, Italy, Germany, China, Belgium, and Bangladesh are among the countries that are the subject of the investigation. China remains the top recipient of international development funds (IDFDI) among developing nations, and it is catching up to the EU in terms of innovation at a three times faster rate than the EU itself. However, there was no discernible correlation between IFDI's substantial negative impact on developing countries like Bangladesh and the sample taken from the economically advanced EU between 2018 and 2022 (Table 9). From a policy standpoint, the findings indicate that traditional and long-term factors such as GDP and interest rates significantly influence IFDI activity more than one-time (Table 12), short-term occurrences and microeconomic factors that affect IFDI behaviour.

## V. CONCLUDING REMARK

The study methodology outlines the difficulties researchers encounter when obtaining reliable data on IFDI across 13 countries. Measurement of IFDI inflow directly impacts entering a country from all sources, including the negative impact of COVID-19 and the Russia-Ukraine war. Tracking the source of FDI funding has also drawn criticism. If the public accepts the regime and its accompanying measures, then doubts about the advantages of an opening IFDI system will have to be dispelled by government and policy markers. Our research has deepened our understanding of the relationship between globalization and poverty and provided insightful policy recommendations for the numerous developing nations that will soon have significant economic influence on the world stage. The study results indicate a substantial adverse effect on IFDI in the selected countries due to the escalation of policy uncertainty on a global scale. The study's conclusions indicated that investment growth in the sample countries as a whole experiences a decrease in IFDI when domestic policy uncertainty exists. Throughout a sample of 13 countries, FDI is influenced by a wide range of variables, including yet not limited to GDP, interest rate, inflation rate, exchange rate, and financial development. Modern communication settings in the global context, privatization and further reforms, port service decentralization, setup of industrial parks, and IFDI continue to flood into Asia despite the region's geopolitical posturing and economic difficulties. Policymakers ought to promote FDI inflows, draw in and support foreign investors, and emphasize the mechanism for mitigating uncertainty by lessening the negative effects of the conflict between Russia and Ukraine during COVID-19.

## VI. REFERENCES

[1] Ahmad, M., Jabeen, G., Irfan, M., Işık, C. and Rehman, A., 2021. Do inward foreign direct investment and economic development improve local environmental quality: aggregation bias puzzle. *Environmental Science and Pollution Research*, 28(26), pp.34676–34696. https://doi.org/10.1007/s11356-021-12734-y.

[2] Alam, Md.K., Tabash, M.I., Billah, M., Kumar, S. and Anagreh, S., 2022. The Impacts of the Russia–Ukraine Invasion on Global Markets and Commodities: A Dynamic Connectedness among G7 and BRIC Markets. *Journal of Risk and Financial Management*, 15(8), p.352. https://doi.org/10.3390/jrfm15080352.

[3] Albulescu, C.T. and Ionescu, A.M., 2018. The long-run impact of monetary policy uncertainty and banking stability on inward FDI in EU countries. *Research in International Business and Finance*, 45, pp.72–81. https://doi.org/10.1016/j.ribaf.2017.07.133.

[4] Alguacil, M., Cuadros, A. and Orts, V., 2008. EU Enlargement and Inward FDI. *Review of Development Economics*, 12(3), pp.594–604. https://doi.org/10.1111/j.1467-9361.2008.00474.x.

[5] Amendolagine, V., Presbitero, A.F., Rabellotti, R. and Sanfilippo, M., 2019. Local sourcing in developing countries: The role of foreign direct investments and global value chains. *World Development*, 113, pp.73–88. https://doi.org/10.1016/j.worlddev.2018.08.010.

[6] Bailey, D., Driffield, N. and Kispeter, E., 2019. Brexit, foreign investment and employment: some implications for industrial policy? *Contemporary Social Science*, 14(2), pp.174–188. https://doi.org/10.1080/21582041.2019.1566563.

[7] Boateng, A., Hua, X., Nisar, S. and Wu, J., 2015. We are examining the determinants of inward FDI: Evidence from Norway. *Economic Modelling*, 47, pp.118–127. https://doi.org/10.1016/j.econmod.2015.02.018.

[8] Borensztein, E., De Gregorio, J. and Lee, J.-W., 1998. How does a foreign direct investment affect economic growth? We are grateful for comments from Robert Barro, Elhanan Helpman, Boyan Jovanovic, Mohsin Khan, Se-Jik Kim, Donald Mathieson, Sergio Rebelo, Jeffrey Sachs, Peter Wickham, and two anonymous referees. Comments by participants in seminars at the 1995 World Congress of the Econometric Society, Korean Macroeconomics Workshop, Kobe University, and Osaka University were constructive. This paper was partially prepared while José de Gregorio and Jong-Wha Lee were at the Research Department, International Monetary Fund. Any opinions expressed are only those of the authors and not those of the institutions with which the authors are affiliated.1. *Journal of International Economics*, 45(1), pp.115–135. https://doi.org/10.1016/S0022-1996(97)00033-0.

[9] Burlea-Schiopoiu, A., Brostescu, S. and Popescu, L., 2023. The impact of foreign direct investment on the economic development of emerging countries of the European Union. *International Journal of Finance & Economics*, 28(2), pp.2148–2177. https://doi.org/10.1002/ijfe.2530.

[10] Charaia, V., Lashkhi, M. and Lashkhi, M., 2022. Foreign Direct Investments during the Coronomic Crisis and Armed Conflict in the Neighbourhood, Case of Georgia. *Globalization and Business*, pp.51–56. https://doi.org/10.35945/gb.2022.13.007.

[11] Chattopadhyay, A.K., Rakshit, D., Chatterjee, P. and Paul, A., 2022. Trends and Determinants of FDI with Implications of COVID-19 in BRICS. *Global Journal of Emerging Market Economies*, 14(1), pp.43–59. https://doi.org/10.1177/09749101211067091.

[12] Chowdhury, A. and Mavrotas, G., 2006. FDI and Growth: What Causes What? *The World Economy*, 29(1), pp.9–19. https://doi.org/10.1111/j.1467-9701.2006.00755.x.

[13] Driffield, N. and Karoglou, M., 2019. Brexit and Foreign Investment in the UK. *Journal of the Royal Statistical Society Series A: Statistics in Society*, 182(2), pp.559–582. https://doi.org/10.1111/rssa.12417.

[14] Grosse, R. and Trevino, L.J., 1996. Foreign Direct Investment in the United States: An Analysis by Country of Origin. *Journal of International Business Studies*, 27(1), pp.139–155. https://doi.org/10.1057/palgrave.jibs.8490129.

[15] Hajian Heidary, M., 2022. A system dynamics model of the impact of COVID-19 pandemic and foreign direct investment in the global supply chain. *Future Business Journal*, 8(1), p.40. https://doi.org/10.1186/s43093-022-00155-3.

[16] Hosen, MS, 2022. The ICT on the education system and it's future prospects in Bangladesh. *AGPE THE ROYAL GONDWANA RESEARCH JOURNAL OF HISTORY, SCIENCE, ECONOMIC, POLITICAL AND SOCIAL SCIENCE*, 3(2), pp.48–60.

[17] Huang, C.-H., Teng, K.-F. and Tsai, P.-L., 2010. Inward and outward foreign direct investment and poverty: East Asia vs. Latin America. *Review of World Economics*, 146(4), pp.763–779. https://doi.org/10.1007/s10290-010-0069-3.

[18] IMF, 2023. *World Economic Outlook (April 2023) - Inflation rate, average consumer prices*. [online] Available at: <https://www.imf.org/external/datamapper/PCPIPCH@WEO> [Accessed 27 August 2023].

[19] Izzeldin, M., Muradoğlu, Y.G., Pappas, V., Petropoulou, A. and Sivaprasad, S., 2023. The impact of the Russian-Ukrainian war on global financial markets. *International Review of Financial Analysis*, 87, p.102598. https://doi.org/10.1016/j.irfa.2023.102598.

[20] Jude, C. and Silaghi, M.I.P., 2016. Employment effects of foreign direct investment: New evidence from Central and Eastern European countries. *International Economics*, 145, pp.32–49. https://doi.org/10.1016/j.inteco.2015.02.003.

[21] Khan, Y.A. and Ahmad, M., 2021. Investigating the impact of renewable energy, international trade, tourism, and foreign direct investment on carbon emission in developing as well as developed countries. *Environmental Science and Pollution Research*, 28(24), pp.31246–31255. https://doi.org/10.1007/s11356-021-12937-3.

[22] Klein, M.W. and Rosengren, E., 1994. The real exchange rate and foreign direct investment in the United States: Relative wealth vs. relative wage effects. *Journal of International Economics*, 36(3), pp.373–389. https://doi.org/10.1016/0022-1996(94)90009-4.

[23] Kottaridi, C., Louloudi, K. and Karkalakos, S., 2019. Human capital, skills and competencies: Varying effects on inward FDI in the EU context. *International Business Review*, 28(2), pp.375–390. https://doi.org/10.1016/j.ibusrev.2018.10.008.

[24] Liadze, I., Macchiarelli, C., Mortimer-Lee, P. and Sanchez Juanino, P., 2023. Economic costs of the Russia-Ukraine war. *The World Economy*, 46(4), pp.874–886. https://doi.org/10.1111/twec.13336.

[25] Moosa, I.A. and Merza, E., 2022. The effect of COVID-19 on foreign direct investment inflows: stylised facts and some explanations. *Future Business Journal*, 8(1), p.20. https://doi.org/10.1186/s43093-022-00129-5.

[26] Nawo, L. and Njangang, H., 2022. The effect of covid-19 outbreak on foreign direct investment: do sovereign wealth funds matter? *Transnational Corporations Review*, 14(1), pp.1–17. https://doi.org/10.1080/19186444.2021.1964313.

[27] Nurmakhanova, M., Elheddad, M., Alfar, A.J.K., Egbulonu, A. and Zoynul Abedin, M., 2023. Does natural resource curse in finance exist in Africa? Evidence from spatial techniques. *Resources Policy*, 80, p.103151. https://doi.org/10.1016/j.resourpol.2022.103151.

[28] OECD, 2023. *Exchange Rate Index*. [online] Available at: <https://data.oecd.org/searchresults/?q=exchange+rate> [Accessed 27 August 2023].

[29] Orhun, E., 2021. The impact of COVID-19 global health crisis on stock markets and understanding the cross-country effects. *Pacific Accounting Review*, 33(1), pp.142–159. https://doi.org/10.1108/PAR-07-2020-0096.

[30] Ramb, F. and Weichenrieder, A.J., 2005. Taxes and the Financial Structure of German Inward FDI. *Review of World Economics*, 141(4), pp.670–692. https://doi.org/10.1007/s10290-005-0051-7.

[31] Strange, R., 2020. The 2020 Covid-19 pandemic and global value chains. *Journal of Industrial and Business Economics*, 47(3), pp.455–465. https://doi.org/10.1007/s40812-020-00162-x.

[32] Teng, Y., Jin, Y., Wen, H., Ye, X. and Liu, C., 2023. Spatial spillover effect of the synergistic development of inward and outward foreign direct investment on ecological well-being performance in China. *Environmental Science and Pollution Research*, 30(16), pp.46547–46561. https://doi.org/10.1007/s11356-023-25617-1.

[33] Uddin, M., Chowdhury, A., Zafar, S., Shafique, S. and Liu, J., 2019. Institutional determinants of inward FDI: Evidence from Pakistan. *International Business Review*, 28(2), pp.344–358. https://doi.org/10.1016/j.ibusrev.2018.10.006.

[34] Udomkerdmongkol, M., Morrissey, O. and Görg, H., 2009. Exchange Rates and Outward Foreign Direct Investment: US FDI in Emerging Economies. *Review of Development Economics*, 13(4), pp.754–764. https://doi.org/10.1111/j.1467-9361.2009.00514.x.

[35] UNCTAD, 2023a. *Global foreign direct investment flows over the last 30 years*. [online] Available at: <https://unctad.org/data-visualization/global-foreign-direct-investment-flows-over-last-30-years> [Accessed 27 August 2023].

[36] UNCTAD, 2023b. *Investment statistics and trends*. [online] Available at: <https://unctad.org/topic/investment/investment-statistics-and-trends> [Accessed 27 August 2023].

[37] WEF, 2022. *Coronavirus (COVID-19): Read all content*. [online] World Economic Forum. Available at: <https://www.weforum.org/focus/coronavirus-covid-19-4236d8b7e9/> [Accessed 12 November 2023].

[38] World Bank, 2023a. *Foreign direct investment, net inflows (BoP, current US$)*. [World Bank Open Data] World Bank Open Data. Available at: <https://data.worldbank.org> [Accessed 27 August 2023].

[39] World Bank, 2023b. *Real interest rate (%) | Data*. [online] Available at: <https://data.worldbank.org/indicator/FR.INR.RINR> [Accessed 27 August 2023].

[40] World Bank, 2023c. *World Bank Maps*. [online] Available at: <https://maps.worldbank.org/projects?status=active> [Accessed 27 August 2023].

[41] WTO, 2023. *News archive: COVID-19 and world trade*. [online] Available at: <https://www.wto.org/english/news_e/archive_e/tfore_arc_e.htm> [Accessed 12 November 2023].
[42] Xia, X. and Liu, W.-H., 2021. China's Investments in Germany and the Impact of the COVID-19 Pandemic. *Intereconomics*, 56(2), pp.113–119. https://doi.org/10.1007/s10272-021-0962-0.
[43] Zaman, K.A.U., 2023. Financing the SDGs: How Bangladesh May Reshape Its Strategies in the Post-COVID Era? *The European Journal of Development Research*, 35(1), pp.51–84. https://doi.org/10.1057/s41287-022-00556-8.